\documentclass[a4paper,11pt]{article}
\pdfoutput=1

\newcommand{\sla}[1]{#1\kern-0.55em/}
\usepackage{jheppub}
\usepackage{graphicx,epsfig}
\usepackage{dcolumn}
\usepackage{bm}
\usepackage{amsmath}
\usepackage{amssymb}
\usepackage{color}
\usepackage{float}
\usepackage{nccmath}

\usepackage[caption = false]{subfig}

\definecolor{grey}{gray}{0.6}

\renewcommand\thefootnote{\fnsymbol{footnote}}

\newcommand{\wimp}{\textsc{WIMP}}

\newcommand{\sigmav}{\ensuremath{\langle \sigma_{\mbox{eff}} v \rangle}}
\newcommand{\gammaphi}{\ensuremath{\Gamma_{\phi}}}
\newcommand{\rhophi}{\ensuremath{\rho_{\phi}}}
\newcommand{\Yphi}{\ensuremath{Y_{\phi}}}
\newcommand{\stdstar}{\ensuremath{\widetilde{\Sigma}^{*}}}
\newcommand{\stdstarmax}{\ensuremath{\widetilde{\Sigma}^{*}_{\rm MAX}}}
\newcommand{\geff}{\ensuremath{g_{eff}}}
\newcommand{\heff}{\ensuremath{h_{eff}}}

\newcommand{\gstar}{\ensuremath{g_{*}^{1/2}}}
\newcommand{\rhotphi}{\ensuremath{\widetilde{\rho}_{\phi}}}

\newcommand{\Oh}{\ensuremath{\Omega h^2}}
\newcommand{\TRH}{\ensuremath{T_{RH}}}
\newcommand{\TRHBBN}{\ensuremath{T_{RH}^{\rm BBN \, lim}}}
\newcommand{\Tfo}{\ensuremath{T_{\rm fo}}}
\newcommand{\Tinit}{\ensuremath{T_{\rm init}}}

\newcommand{\Log}{\ensuremath{{\rm log}}}

\begin{document}

\hfill {\tt  CERN-TH-2018-150, KCL-PH-TH/2018-30}  

\def\thefootnote{\fnsymbol{footnote}}
 
\begin{center}

\vspace{2.cm}

{\Large\bf Dark Matter Casts Light on the Early Universe}

\setlength{\textwidth}{11cm}
                    
\vspace{1cm}
{\large\bf  
A.~Arbey$^{\,a,b,}$\footnote{Also Institut Universitaire de France, 103 boulevard Saint-Michel, 75005 Paris, France}$^{,}$\footnote{Email: alexandre.arbey@ens-lyon.fr}, 
J.~Ellis$^{b,c,d,}$\footnote{Email: John.Ellis@cern.ch},
F.~Mahmoudi$^{a,b,*,}$\footnote{Email: nazila@cern.ch},
G.~Robbins$^{a,e}$\footnote{Email: glenn.robbins@univ-lyon1.fr}
}
 
\vspace{0.5cm}
{\em $^a$Univ. Lyon, Univ. Lyon 1, CNRS/IN2P3, Institut de Physique Nucl\'eaire de Lyon,\\ UMR5822, F-69622 Villeurbanne, France}\\[0.2cm]
{\em $^b$Theoretical Physics Department, CERN, CH-1211 Geneva 23, Switzerland} \\[0.2cm]
{\em $^c$Theoretical Particle Physics and Cosmology Group, Department of Physics,\\
King's College London, London WC2R 2LS, United Kingdom}\\[0.2cm] 
{\em $^d$National Institute of Chemical Physics \& Biophysics, R{}\"avala 10, 10143 Tallinn, Estonia}\\[0.2cm]
{\em $^e$Univ. Lyon, Univ. Lyon 1, ENS de Lyon, CNRS, Centre de Recherche Astrophysique de Lyon UMR5574, F-69230 Saint-Genis-Laval, France}

\end{center}

\renewcommand{\thefootnote}{\arabic{footnote}}
\setcounter{footnote}{0}

\vspace{0.5cm}
\thispagestyle{empty}
\centerline{\bf ABSTRACT}
\vspace{0.5cm}
We show how knowledge of the cold dark matter (CDM) density can be used, in conjunction with
measurements of the parameters of a scenario for beyond the Standard Model (BSM) physics, to
provide information about the evolution of the Universe before Big Bang Nucleosynthesis (BBN).
As examples of non-standard evolution, we consider models with a scalar field that may decay into
BSM particles, and quintessence models. We illustrate our calculations using various
supersymmetric models as representatives of classes of BSM scenarios in which the CDM density is 
either larger or smaller than the observed density when the early Universe is assumed to be radiation-dominated. 
In the case of a decaying scalar field, we show how the CDM density can constrain the initial scalar
density and the reheating temperature after it decays in BSM scenarios that would yield overdense
dark matter in standard radiation-dominated cosmology, and how the decays of the scalar field into
BSM particles can be constrained in scenarios that would otherwise yield underdense CDM.
We also show how the early evolution of the quintessence field can be constrained in BSM
scenarios.


\newpage

\section{Introduction}
\label{sec:intro}
 
The very early Universe before Big-Bang Nucleosynthesis (BBN) is a little-known cosmological era that should provide the answers to several very important questions, such as the origin of the baryon asymmetry in the Universe -- possibly due to leptogenesis, the nature of the electroweak and perhaps other phase transitions, the possibility of grand unification, the mechanism for inflation, etc.. Unfortunately, as of today we have no direct observations of the period before recombination at $\sim 1$~eV, though some constraints can be set using the abundances of the elements generated during BBN, and the cosmic microwave background (CMB) constrains models of inflation. High-energy colliders such as the Large Hadron Collider (LHC) can probe the state of matter at energies $\sim$~GeV and particle interactions at energies $\sim$ TeV, but the other properties of the early Universe, such as its expansion rate, are still relatively unconstrained.

In this paper we propose to use understanding of the properties of relic dark matter (DM) particles obtained from particle physics to obtain constraints on the properties of the very early Universe at temperatures $\sim 10-100$ GeV, orders of magnitude above the scale of BBN.

For this purpose, we consider an observable linking particle physics and cosmology, namely the DM relic density. 
We assume that DM is cold, and composed of some type of stable weakly-interacting massive particle (WIMP)
that was in thermal equilibrium in the early Universe and subsequently froze out. 
The cold dark matter density has been measured very precisely by the Planck Collaboration using the CMB 
and observations of the more recent Universe \cite{Ade:2015xua}. The Standard Model (SM) of particle physics 
does not provide any cold dark matter candidate, but many new physics beyond the SM (BSM) have such candidates. 
The dark matter relic density can be computed in any given BSM scenario, {\it under the assumption} that the early Universe
was dominated by (SM) radiation, and very strict constraints can be set on the parameters of the BSM scenario using the Planck measurements \cite{Jungman:1995df}.

The hypothesis of radiation domination in the early Universe is correct at temperatures below $\sim$ MeV, 
as indicated by the constraints from BBN and the CMB \cite{Olive:1999ij,Barger:2003zg,Fields:2014uja}. However, it is possible that it does not hold at 
higher temperatures. In particular, many cosmological scenarios, such as late inflation~\cite{Wetterich:1994bg,Ferreira:1997hj,Capozziello:2005tf}, 
dark energy~\cite{Ratra:1987rm,Zlatev:1998tr,Amendola:1999er,Caldwell:1999ew,Chiba:1999ka,Bento:2002ps,Nojiri:2005pu,Tsujikawa:2013fta}, a dark fluid~\cite{Peebles:2000yy,Bilic:2001cg,Arbey:2006it}, Higgs inflation \cite{Bezrukov:2007ep,Bezrukov:2010jz}, late-decaying moduli~\cite{Dine:1995uk,Banks:1995dt,Moroi:2001ct,Nakamura:2006uc}, dilatons \cite{deCarlos:1993wie,Gasperini:1994xg}, etc., invoke cosmological scalar fields that may have co-existed with radiation at temperatures $\sim$ GeV or TeV. Several studies (see, for example,~\cite{Kamionkowski:1990ni,Salati:2002md,Rosati:2003yw,Comelli:2003cv,Pallis:2004yy,Pallis:2005hm,Gelmini:2006pw,Gelmini:2006pq,Arbey:2008kv,Arbey:2009gt,Lola:2009at,Drees:2017iod, DEramo:2017gpl}) have shown that such scalar fields could have altered the relic density.

In any given BSM scenario, a deviation of the measured cold dark matter density from a calculation based on
measurements of the model parameters and standard radiation-dominated expansion would be a signature of novel phenomena in the very early Universe. One might argue that, if the calculated relic density is different from the measured dark matter density, the corresponding BSM scenario is disfavoured. 
Here, however, we propose to reverse this argument: if the calculated relic density is different from the 
measured dark matter density, it could be because of novel phenomena in the early Universe. 
This orthogonal point of view will become particularly important if new particles are discovered at colliders 
or in dark matter detection experiments: using dark matter observables, it is not possible to constrain
BSM scenarios in isolation, but the constraints have to be applied simultaneously to a combination of BSM and cosmological scenarios.

For this analysis, we study two different realistic cosmological scenarios: the case of a decaying scalar field, e.g., a modulus field, which modifies the energy content of the Universe and also injects entropy or BSM particles, and the case of a quintessence field, which could modify the energy content on its way to fulfilling its original purpose of generating dark energy with negative pressure in the recent Universe. 

The rest of this paper is organised as follows. In Section 2 we review the standard calculation of relic density. 
Then, in Section 3 we introduce cosmological scalar field scenarios that can impact the relic density calculation, 
and discuss their possible effects. Next, in Section 4 we introduce as illustrations of BSM scenarios
a selection of supersymmetric scenarios where the measured relic density can differ from that
calculated assuming radiation-dominated expansion. 
Our results are given in Section 5 and our conclusions in Section 6.

\section{Relic Density Calculation}
The relic density calculation is generally performed in the standard cosmological model,
in which the expansion rate of the Universe is given by the Friedmann equation. In the early Universe when the radiation density dominates
this reduces to:
\begin{equation}
H^2=\left(\frac{\dot{a}}{a}\right)^2=\frac{8 \pi G}{3} \rho_{rad} \ ,
\end{equation}
where $a$ is the cosmological scale factor and $H$ the Hubble parameter. The radiation density reads
\begin{equation}
\rho_{rad}(T)=g_{\mbox{eff}}(T) \frac{\pi^2}{30} T^4 \ ,
\end{equation}
where $g_{\mbox{eff}}$ is the effective number of degrees of freedom of radiation, which is given by the 
particle content of the Standard Model and the QCD equation of state 
(see, for example, \cite{Hindmarsh:2005ix,Drees:2015exa}).

Assuming that, in a given BSM scenario, only the lightest BSM particle is stable, and constitutes a
suitable dark matter candidate that was originally in thermal equilibrium, the number of relic particles is 
obtained by solving the Boltzmann evolution equation~\cite{Gondolo:1990dk,Edsjo:1997bg}:
\begin{equation}
dn/dt=-3Hn-\langle \sigma_{\mbox{eff}} v\rangle (n^2 - n_{\mbox{eq}}^2) \ ,
\end{equation}
where $n$ is the number density of BSM particles, $n_{\mbox{eq}}$ is their equilibrium density, 
and $\langle \sigma_{\mbox{eff}} v\rangle$ is the thermal average of the annihilation rate of 
pairs of BSM particles to SM particles.

To define $\langle \sigma_{\rm{eff}}v \rangle$, it is useful to define first the annihilation rate of BSM particles $i$ and $j$ into SM particles $k$ and $l$:
\begin{equation}
W_{ij\to kl} = \frac{p_{kl}}{16\pi^2 g_i g_j S_{kl} \sqrt{s}} \sum_{\rm{internal~d.o.f.}} \int \left| \mathcal{M}(ij\to kl) \right|^2 d\Omega  \ ,
\end{equation}
where $\mathcal{M}$ is the transition amplitude, $s$ is the centre-of-mass energy squared, 
$g_i$ is the number of degrees of freedom of the particle $i$, $p_{kl}$ is the final centre-of-mass momentum, given by
\begin{equation}
p_{kl} = \frac{\left[s-(m_k+m_l)^2\right]^{1/2} \left[s-(m_k-m_l)^2\right]^{1/2}}{2\sqrt{s}} \ ,
\end{equation}
and $S_{kl}$ is a symmetry factor equal to 2 for identical final particles and to 1 otherwise.

The thermal average of the effective cross section is given by:
\begin{equation}
\langle \sigma_{\rm{eff}}v \rangle = \dfrac{\displaystyle\int_0^\infty dp_{\rm{eff}} p_{\rm{eff}}^2 W_{\rm{eff}}(\sqrt{s}) K_1 \left(\dfrac{\sqrt{s}}{T} \right) } { m_{relic}^4 T \left[ \displaystyle\sum_i \dfrac{g_i}{g_{LSP}} \dfrac{m_i^2}{m_1^2} K_2 \left(\dfrac{m_i}{T}\right) \right]^2} \ ,
\end{equation}
where $K_1$ and $K_2$ are the modified Bessel functions of the second kind of order 1 and 2 respectively, 
and $W_{\rm eff}$ is an effective annihilation rate:
\begin{equation}
W_{\rm{eff}} \equiv \frac{1}{g_{relic}^2 p_{\rm{eff}}} \sum_{ij} g_i g_j p_{ij} W_{ij} \ ,
\end{equation}
with
\begin{equation}
p_{\rm{eff}}(\sqrt{s}) = \frac{1}{2} \sqrt{(\sqrt{s})^2 -4 m_{relic}^2} \ ,
\end{equation}

In order to solve the Boltzmann equation, it is necessary to have a link between time and temperature, 
which is given under the assumption of adiabaticity by
\begin{equation}
 \frac{ds_{rad}}{dt} = - 3 H s_{rad} \ ,
\end{equation}
where the radiation entropy density is given by
\begin{equation}
s(T)=h_{\mbox{eff}}(T) \frac{2 \pi^2}{45} T^3 \ ,
\end{equation}
with $h_{\mbox{eff}}$ the effective number of entropic degrees of freedom of radiation. 

To solve this set of equations, one defines the ratio of the number density of BSM particles to the radiation entropy density $Y(T) \equiv n(T)/s_{rad}(T)$, 
and the ratio of the relic particle mass to the temperature, $x \equiv m_{relic}/T$, and combines them into~\cite{Gondolo:1990dk,Edsjo:1997bg}:
\begin{equation}
\frac{dY}{dx}=-\sqrt{\frac{\pi}{45 G}}\frac{g_*^{1/2} m_{relic}}{x^2} \langle \sigma_{\mbox{eff}} v\rangle (Y^2 - Y^2_{\mbox{eq}})  \ , \label{main}
\end{equation}
with
\begin{equation}
g_*^{1/2}=\frac{h_{\mbox{eff}}}{\sqrt{g_{\mbox{eff}}}}\left(1+\frac{T}{3 h_{\mbox{eff}}}\frac{dh_{\mbox{eff}}}{dT}\right)  \ .
\end{equation}
The freeze-out temperature $T_f$ is the temperature at which the relic particle leaves the initial thermal equilibrium, 
which is expected to happen at $\sim m_{relic} / 10 ~ \sim 10-100$ GeV in many BSM WIMP scenarios.

Solving the equations down to the present temperature $T_0$, {we find that $Y$ approaches a constant asymptotic value} and the relic density so obtained is~\cite{Gondolo:1990dk,Edsjo:1997bg}:
\begin{equation}
\Omega_{relic} h^2 = \frac{m_{relic} s(T_0) Y(T_0) h^2}{\rho_c^0} \approx 2.755\times 10^8 \frac{m_{relic}}{1 \mbox{ GeV}} Y(T_0) \ ,
\end{equation}
where $\rho_c^0$ is the critical density of the Universe, given by
\begin{equation}
H^2_0 = \frac{8 \pi G}{3} \rho_c^0  \ ,
\end{equation}
and $H_0$ is the Hubble constant.
The relic density can then be compared to the measurements of the dark matter density by the Planck Collaboration \cite{Ade:2015xua} 
to set constraints on the BSM scenarios. 

In the following, we use {\tt SuperIso~Relic~v4.0}~\cite{Arbey:2009gu,Arbey:2011zz,Arbey:2018msw} to compute the relic density. 
Since it was shown that the theoretical uncertainties due to the cross section calculation at tree level and to the uncertainties
in the QCD equation of state are of the order of a tenth~\cite{Hindmarsh:2005ix,Baro:2007em,Arbey:2008kv,Arbey:2009gt,Harz:2012fz,Drees:2015exa}, 
we add a 10\% theoretical error to the Planck measurements and obtain the following 95\% C.L. interval:
\begin{equation} 
0.095<\Oh<0.1428\,.
\end{equation}

\section{Cosmological Scenarios}
The standard relic density calculation can be modified by the presence of scalar fields in the early Universe, 
which can affect the expansion rate by adding a new energy density, generate non-thermal relic particles,
or inject entropy and affect the relation between time and temperature. In the following, we consider the case of a decaying pressureless scalar field and of quintessence as realistic examples of cosmological models affecting the early Universe. 
Since the freeze-out occurs at $\sim 10-100$~GeV, a large deviation from the standard model of cosmology at this temperature could modify strongly the results, without having other consequences for the observable Universe. The strongest constraints that can be set on such cosmological scenarios are those from BBN. In the following, we compute BBN constraints for the scenarios of interest using {\tt AlterBBN~v2.0}~\cite{Arbey:2011nf,Arbey:2018zfh} and the conservative limits on the abundances of the elements given in~\cite{Jedamzik:2006xz}.

\subsection{Decaying primordial scalar field}

We consider a pressureless scalar field $\phi$ of mass $m_{\phi}$ that decays into radiation with a 
width $\Gamma_{\phi}$, and into BSM particles with a branching ratio $b$ \cite{Gelmini:2006pw,Gelmini:2006pq}. 
The evolution in time of the scalar field density $\rho_{\phi}$ and the \wimp\ density $n=\rho_{\chi}/m_{\chi}$ can be determined from the following equations:
\begin{align}
\frac{d\rhophi}{dt}&\;=\;-3H \rhophi -\gammaphi \rhophi \ ,\\
    \frac{dn}{dt}&\;=\; -3Hn -\sigmav \left(n^{2}- n_{eq}^{2} \right)+ \frac{b}{m_{\phi}} \gammaphi \rhophi \ ,
\end{align}
where \sigmav\ is the thermally-averaged \wimp\ annihilation cross section, $n_{eq}$ is the \wimp\ equilibrium density, and $H$ is the Hubble {parameter}, which depends on the total energy density in the Universe:
\begin{equation}
  H^{2}\;=\; \frac{8\pi}{3 M_{p}^{2}} \left( \rho_{\phi}+ \rho_{rad} + \rho_{\chi} \right) \ .
\end{equation}

We assume that the thermalisation of the decay products of the scalar field occurs instantaneously\footnote{Discussions of the effect of other thermalisation assumptions can be found in \cite{Allahverdi:2002pu,Mukaida:2015ria}.}.
In order to obtain a relation between the time and the temperature, one may use the following equation for the evolution of the radiation entropy density \cite{Kolb:1990vq}:
\begin{equation}
\frac{ds_{rad}}{dt}\;=\;-3Hs_{rad} +\frac{\gammaphi \rhophi}{T}\;=\;-3H \left(1-\stdstar \right)s_{rad} \ , \label{eq:entropy_derivative}
\end{equation}
with 
\begin{equation}
\stdstar\; \equiv \; \frac{\gammaphi \rhophi}{3H\, T\,s_{rad}}\ .
\label{eq:stdstarev}
\end{equation}
The energy and entropy densities of radiation can be determined from the temperature according to:
\begin{align}
  \left\{
  \begin{aligned}
    \rho_{rad}&\;=\; \frac{\pi^{2}}{30} \geff(T) T^{4} \ ,\\
 	s_{rad}&\;=\; \frac{2\pi^{2}}{45} \heff(T) T^{3} \ ,
 \end{aligned}
  \right.
\end{align}
where \geff\ and \heff\ are the number of degrees of freedom of radiation energy and the entropy, respectively. We use the QCD equation of state ``B'' of Ref.~\cite{Hindmarsh:2005ix} in our analysis.

The decay width may conveniently be expressed as a function of the reheating temperature \TRH\ \cite{Gelmini:2006pw,Gelmini:2006pq}, which is the temperature at which the scalar field density starts to be significantly reduced:
\begin{equation}
    \Gamma_{\phi}\;=\; \sqrt{\frac{4 \pi^{3} \geff(\TRH)}{45}}\; \frac{\TRH^{2}}{M_{p}} \ .
  \label{eq:stdstar}
\end{equation}
We also define $\tilde\rho_\phi \equiv \rho_\phi/\rho_{rad}$ and the initial condition $\kappa_\phi \equiv \rho_\phi (T_{init})/\rho_\gamma (T_{init})$.

In the following we assume that the period of interest for the relic density occurs when the radiation entropy density decreases with time, which corresponds to $\stdstar<1$. This imposes a maximal temperature $T_{\rm max}$ for the validity of the following discussion, which corresponds to the temperature at which $\stdstar=1$. The above equations can be re-written as derivatives of $\Yphi=\rho_{\phi}/s_{\rm rad}$ and $Y=n/s_{\rm rad}$:
\begin{align}
\frac{d \Yphi}{dx}&=- \frac{\alpha_0}{x} \frac{\stdstar}{1-\stdstar} \left(\Yphi +\frac{m_{\chi}}{x}  \right)
\label{eq:dYphi} \ ,\\
\frac{dY}{dx}=&- \frac{\alpha_0}{x} \frac{s_{\rm rad}}{1-\stdstar}  \frac{1}{3H}\sigmav \left(Y^2 -Y_{eq}^2 \right)-\frac{\alpha_0}{x} \frac{\stdstar}{1-\stdstar} 
\left( Y - \frac{b}{m_{\phi}} \frac{m_{\chi}}{x}  \right) \ ,\label{eq:dY}
\end{align}
with
\begin{equation}
\alpha_0=\frac{3 \gstar \geff^{1/2}}{\heff} \approx 3 \ ,
\end{equation}
where $x=m_{\chi}/T$.\\

Eqs. (\ref{eq:dYphi}) and (\ref{eq:dY}) are controlled by the parameter \stdstar\ defined in Eq. (\ref{eq:stdstarev}).
In order to understand its role, we consider the entropy time-derivative equation (\ref{eq:entropy_derivative}) in the case where \stdstar{} is constant.
If $T \propto t^\alpha$ and the scale factor $a \propto t^\beta$, then $H=\beta t^{-1}$ and we obtain:
\begin{equation}
3 \alpha =- 3 \beta (1-\stdstar) \ .
\end{equation}
Thus, $\beta= -\alpha/(1- \stdstar)$ and $a \propto t^{-\alpha/ (1- \stdstar)} \propto T^{-1/(1-\stdstar)}$. 
After freeze-out, the \wimp\ density verifies $\rho_{\chi} \propto a^{-3}$, so $\rho_{\chi} \propto T^{3/(1- \stdstar)}$.
The \wimp\ density will therefore be diluted very fast as \stdstar\ $\to 1$. 
In fact, one can derive a maximum value for \stdstar\ where $d\log (\stdstar) / d \log (x)=0$.
In the limit $\rho_{\phi}\gg \rho_{\rm rad}$, $\stdstar \propto x^{5/2}\Yphi^{1/2}$ according to Eq.~(\ref{eq:stdstar}).
Thus the maximum value of \stdstar\ is reached when $d \log (\Yphi) / d \log (x)=-5$. Using Eq.~(\ref{eq:dYphi}) we obtain the condition
\begin{equation}
- \alpha_0 \frac{\stdstarmax}{1-\stdstarmax} \left(1+ \frac{T}{\Yphi} \right) = \frac{d \log (\Yphi)}{d \log (x)} = -5 \ ,
\end{equation}
from which it follows that
\begin{equation}
\frac{\stdstarmax}{1-\stdstarmax}= \frac{5}{\alpha_0}\frac{1}{1+ \frac{T}{\Yphi}}<\frac{5}{\alpha_0} \lesssim 1.66 \ ,
\end{equation}
which leads to 
\begin{equation}
\stdstar<\frac{5}{\alpha_0} \frac{1}{1+\frac{5}{\alpha_0}} \approx 0.625 \ .
\end{equation}
This prevents any singularities in the term $\stdstar / 1-\stdstar$, but limits the strength of the dilution.

We have seen that the scalar field density can decrease in two ways: either by decay, or by dilution. Thus,
the presence of the scalar field may modify the \wimp\ relic density from that calculated 
in the standard model of cosmology in three different ways. First, {\wimp}s can be diluted in the same way as the scalar field. As this phenomenon only changes the evolution of the temperature with time, 
it does not affect the \wimp\ density at a given temperature during thermal equilibrium, 
since the equilibrium density is determined by the temperature alone.
Secondly, if the scalar field decays into BSM particles, the \wimp\ density may increase. 
If the decay happens before freeze-out, however, the decay products will annihilate and there would be no consequence on the relic density.

Thirdly, if the scalar field density is large enough, it will change significantly the Hubble {parameter} 
and the freeze-out will occur sooner, thus increasing the density at freeze-out compared to the standard calculation. 
However, as we shall see, this last case corresponds also to that where dilution is important.
Therefore, the only way to increase the relic density is if the scalar field decays {also into BSM} particles.

\subsection{Quintessence}

As an alternative, we also consider a quintessence field\footnote{See, for example, \cite{Tsujikawa:2013fta} for a review of quintessence models.},
which satisfies the continuity equation: 
\begin{equation}
 \frac{d\rhophi}{dt}\;=\; 3H (\rhophi + P_{\phi}) \ ,
\end{equation}
where the pressure and the energy density of the scalar field are $P_{\phi}=\dot{\phi}^2/2 - V(\phi)$ and $\rhophi=\dot{\phi}^2/2 + V(\phi)$, respectively. 

\begin{figure}[t!]
\centering
\includegraphics[width=0.5\columnwidth]{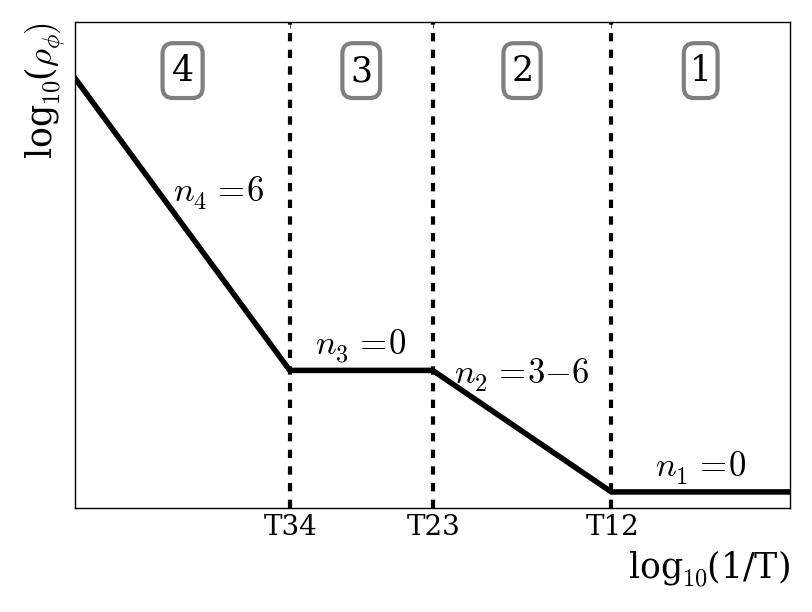}
\caption{\it Evolution with temperature of the scalar field density in representative power-law models of quintessence.\label{fig:model}}
\end{figure}

We have computed the scalar field density evolution with the temperature for three different standard quintessence potentials $V(\phi)$~\cite{Tsujikawa:2013fta}: a double exponential \cite{Barreiro:1999zs}, an inverse power law \cite{Ferreira:1997hj}, and a pseudo-Nambu-Goldstone boson potential \cite{Frieman:1995pm}. 
We find that the scalar field density can be well approximated for the three potentials 
with a power law of slope 6 at high temperatures (zone 4 of Figure~\ref{fig:model}) and of slope 0 at low temperatures coinciding with the measured dark energy density (zone 1). 
In the case of the double exponential potential, two additional power-law changes occur: the first
to a slope 0 (zone 3) and then to a slope ranging from 3 to 6 (zone 2).
Hence, we consider a simplified model whose free parameters are the temperatures $T_{34}$, $T_{23}$, 
$T_{12}$ at which the power-law changes occur, together with the slope in zone 2, $n_2$. 

In this model, there is no way to reduce the relic density compared to the standard cosmological model. 
The only possible influence of the scalar field is the \wimp\ density at freeze-out. 
If the scalar field density is large enough while the \wimp\ is in thermal equilibrium,
the Hubble {parameter} can be enhanced compared to the standard cosmological model. 
This would have the effect of advancing freeze-out and thereby increasing the relic \wimp\ density.

\section{New Physics Scenarios}
In order to illustrate the possible implications of such cosmological scenarios, we consider variants of the minimal supersymmetric extension of the Standard Model (MSSM) with CP and $R$-parity conservation, 
which is representative of a large class of WIMP models. The lightest neutralino is a well-motivated candidate for dark matter \cite{Jungman:1995df}, and we assume in the following that 100\% of cold dark matter is composed of neutralinos. The neutralino can be bino-like, wino-like, higgsino-like or a mixed state. These candidates are weakly-interacting, and in conventional calculations bino-like neutralinos 
have in general a too large a relic density, apart in cases where they are associated with near-degenerate supersymmetric particles with which they can coannihilate, or if annihilations are enhanced by resonances such as heavy Higgs bosons. 
Winos and Higgsinos can reach a relic density close to the observed dark matter abundance via
coannihilations with charginos and/or neutralinos that are nearly degenerate with the lightest neutralino.
On the other hand, light winos and Higgsinos generally have too small a relic density.

In the following we first choose as specific examples one MSSM scenario which would yield overdense DM according to the standard cosmological calculation, and one that would yield underdense DM. We also consider a sample of points in the phenomenological MSSM (pMSSM) with 19 free parameters specified at a low energy scale (the pMSSM19).

\subsection{Benchmark Point A}

We first consider a point with a relic density that would be too large (Point A) according to the standard cosmological calculation. 
For this we modify the parameters of the best-fit point of the pMSSM with 11 free parameters
specified at a low energy scale (the pMSSM11), which was found in~\cite{Bagnaschi:2017tru} taking into account the constraints from $\sim36$ fb$^{-1}$ of  LHC data at 13 TeV, including those from direct searches for supersymmetric (SUSY) particles at the LHC, measurements of  the Higgs boson mass and signal strengths, 
LHC searches for the heavier MSSM Higgs bosons, precision electroweak  observables, the measurement of $(g-2)_\mu$ \cite{Bennett:2006fi}, and flavour physics constraints from $B$- and $K$-physics observables. In addition, the constraints from the direct dark matter detection experiments PICO60 \cite{Amole:2017dex}, XENON1T \cite{Aprile:2017iyp} and PandaX-II \cite{Cui:2017nnn} were taken into account,
together with the previous accelerator and astrophysical measurements. The cosmological constraint on the 
cold dark matter density measured by Planck \cite{Ade:2015xua} was also considered. The relic density at this point is therefore close to the measured dark matter density, but it is possible to increase the relic density while respecting the other constraints.
This point has a bino-like neutralino of mass 381 GeV. As commented above, binos tend to have a relic density that is too large. However, thanks to the small mass splittings with the sleptons of the first and second generations, the relic density of this points is very close to the measured dark matter density.
In order to obtain a larger relic density, we increase the mass parameter $M_{\tilde{l}_{1,2}}$ of the sleptons of first and second generation, taking $M_{\tilde{l}_{1,2}}=400$ GeV. The mass of the lightest neutralino is 381 GeV and the next-to-lightest supersymmetric particles are the right-handed selectron and smuon of mass 423 GeV. The mass splitting is large enough so that the impact of the co-annihilations is limited. We obtain a relic density $\Oh= 1.27$ according to the standard cosmological calculation, and a freeze-out temperature $\Tfo \approx 16$ GeV. The parameters of Point A are given in Table~\ref{CMSSMA} and the spectrum is generated with {\tt SOFTSUSY}~\cite{Allanach:2001kg}.
\begin{table}[t!]
\centering
\begin{tabular}{cccccc}
\hline
$M_1$& $M_2$& $M_3$& $\mu$&$M_{A^0}$& $\tan \beta$ \\
\hline
-391 &1240 &-1714 &-5739 &4221 & 18.8 \\
\hline
\end{tabular}
\vspace{0.25cm}
\\
\begin{tabular}{ccccc}
\hline
$M_{\tilde{q}_{1,2}}$&$M_{\tilde{q}_{3}}$& $M_{\tilde{l}_{1,2}}$& $M_{\tilde{l}_{3}}$& $A_0$ \\
\hline
1996 & 4058 & 400 & 1365 & 5372 \\
\hline
\end{tabular}
\caption{\it The pMSSM11 parameter values (in GeV) of Point A.\label{CMSSMA}}
\end{table}

\subsection{Benchmark Point B}

In this case we modify the best-fit point in the constrained MSSM (CMSSM) found in~\cite{Bagnaschi:2017tru}.
This point has a higgsino-like neutralino and a relic density close to the dark matter density measured by Planck. 
We decrease $M_{12}$ to $3872$ GeV in order to get a lower value of the relic density: $\Omega h^2=5.907\times 10^{-3}$ and use {\tt SOFTSUSY}~\cite{Allanach:2001kg} to calculate the spectrum.
The parameters of point B are given in Table~\ref{CMSSMB}.

\begin{table}[h!]
\centering
\begin{tabular}{ccccc}
\hline
$M_0$& $M_{12}$ & $\tan \beta$ & $A_0$& {\rm sign}($\mu$)\\
\hline
10931 & 3872 & 52.9 & 9188 &$+1$\\
\end{tabular}
\caption{\it The CMSSM parameter values {(in GeV when applicable)} of Point B.\label{CMSSMB}}
\end{table}

\subsection{Sample of pMSSM19 Points}

We consider in addition a sample of points in the pMSSM19 generated using {\tt SOFTSUSY}~\cite{Allanach:2001kg} with a flat random sampling over the ranges given in Table~\ref{tab:pmssm} for the 19 parameters. After checking the theoretical validity of each point, we require it to have a light Higgs boson with mass between 122 and 128 GeV. 
We also require the lightest neutralino to be the lightest supersymmetric particle that constitutes dark matter, using the set-up presented in~\cite{Arbey:2011un,Arbey:2011aa,Arbey:2017eos}. As the neutralino can be bino-like, wino-like, Higgsino-like or a mixed state, this approach allows considerable flexibility, making our analysis sufficiently general that it can indicate the possibilities also in other dark matter models. 

\begin{table}[t!]
\begin{center}
\begin{tabular}{|c|c|}
\hline
~~~~Parameter~~~~ & ~~~~Range (in GeV)~~~~ \\
\hline\hline
$M_A$ & [50, 2000] \\
\hline
$M_1$ & [-3000, 3000] \\
\hline
$M_2$ & [-3000, 3000] \\
\hline
$M_3$ & [50, 3000] \\
\hline
$A_d=A_s=A_b$ & [-10000, 10000] \\
\hline
$A_u=A_c=A_t$ & [-10000, 10000] \\
\hline
$A_e=A_\mu=A_\tau$ & [-10000, 10000] \\
\hline
$\mu$ & [-3000, 3000] \\
\hline
$M_{\tilde{e}_L}=M_{\tilde{\mu}_L}$ & [0, 3000] \\
\hline
$M_{\tilde{e}_R}=M_{\tilde{\mu}_R}$ & [0, 3000] \\
\hline
$M_{\tilde{\tau}_L}$ & [0, 3000] \\
\hline
$M_{\tilde{\tau}_R}$ & [0, 3000] \\
\hline
$M_{\tilde{q}_{1L}}=M_{\tilde{q}_{2L}}$ & [0, 3000] \\
\hline
$M_{\tilde{q}_{3L}}$ & [0, 3000] \\
\hline
$M_{\tilde{u}_R}=M_{\tilde{c}_R}$ & [0, 3000] \\
\hline
$M_{\tilde{t}_R}$ & [0, 3000] \\
\hline
$M_{\tilde{d}_R}=M_{\tilde{s}_R}$ & [0, 3000] \\
\hline
$M_{\tilde{b}_R}$ & [0, 3000] \\
\hline
$\tan\beta$ & [1, 60]\\
\hline
\end{tabular}
\caption{\it The pMSSM19 parameter ranges used in our scan.\label{tab:pmssm}}
\end{center}
\end{table}

\section{Results}
\subsection{Decaying primordial scalar field}

\noindent We consider first the cosmological scenario with a scalar field decaying into radiation and SUSY particles. 
{We perform} a scan over the reheating temperature \TRH\ and the initial scalar field density parametrised as the ratio between the scalar field density and the photon density at $T=\Tinit$ , $\kappa_\phi=\frac{\rho_\phi}{\rho_\gamma}(T=\Tinit)$, and calculate the relic density of Points A and B specified in Section 4. We consider different values of the parameter $\displaystyle\eta=b \left(\frac{\text{1 GeV}}{m_{\phi}} \right)$, in order to study the effect of non-thermal production of SUSY particles on the relic density. In each case we derive constraints on the scalar field parameters for our sample of pMSSM19 points so as to investigate the influence of the neutralino properties on the limits derived from the relic DM density.

We start integrating the Boltzmann equations at a temperature $\Tinit=40$ GeV for point A and $\Tinit=20$ GeV for point B. 
For our sample of pMSSM19 points, we use $\Tinit=1.5 \times \Tfo$, where $\Tfo$ is the freeze-out temperature in the standard cosmological model. These choices were made in order to reduce the computation time while starting the calculation sufficiently long before freeze-out and the decay of the scalar field.

\subsubsection{Point with a large relic density}
\begin{figure}
\begin{center}
\includegraphics[width=0.6\columnwidth]{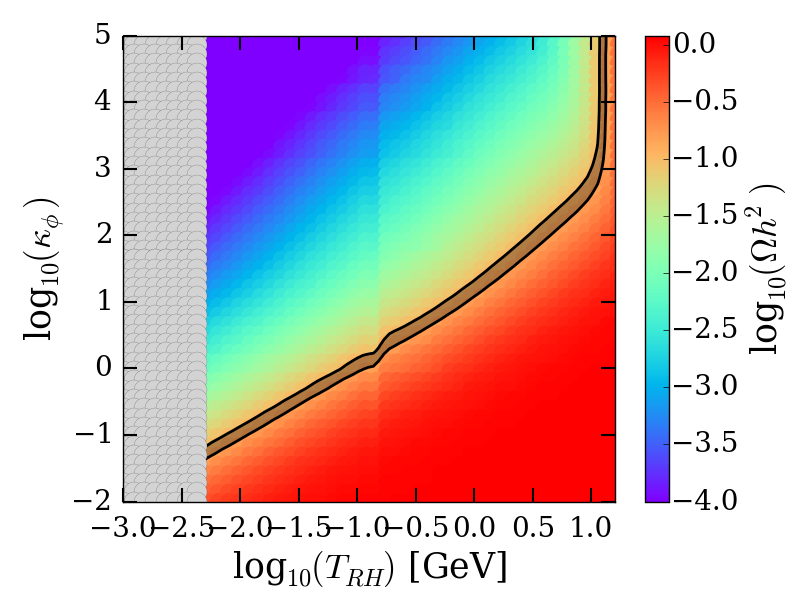}
\caption{\it The relic density $\Log_{10}(\Oh)$ of Point A, indicated by the colour code in the legend, as a function {of} \TRH\ and $\kappa_\phi$. 
Parameter sets consistent with the Planck constraints lie along the darker shaded strip. 
The grey zone at small \TRH\ is excluded by BBN constraints.\label{fig:pmssm11_Tfo>TRH}}
\end{center}
\end{figure}

\begin{figure}
\subfloat[$T_{RH}=0.01$ GeV, $\kappa_\phi=100$, $\Tinit=40$ GeV]{\hspace*{-1.cm}\includegraphics[width=0.46\columnwidth\qquad]{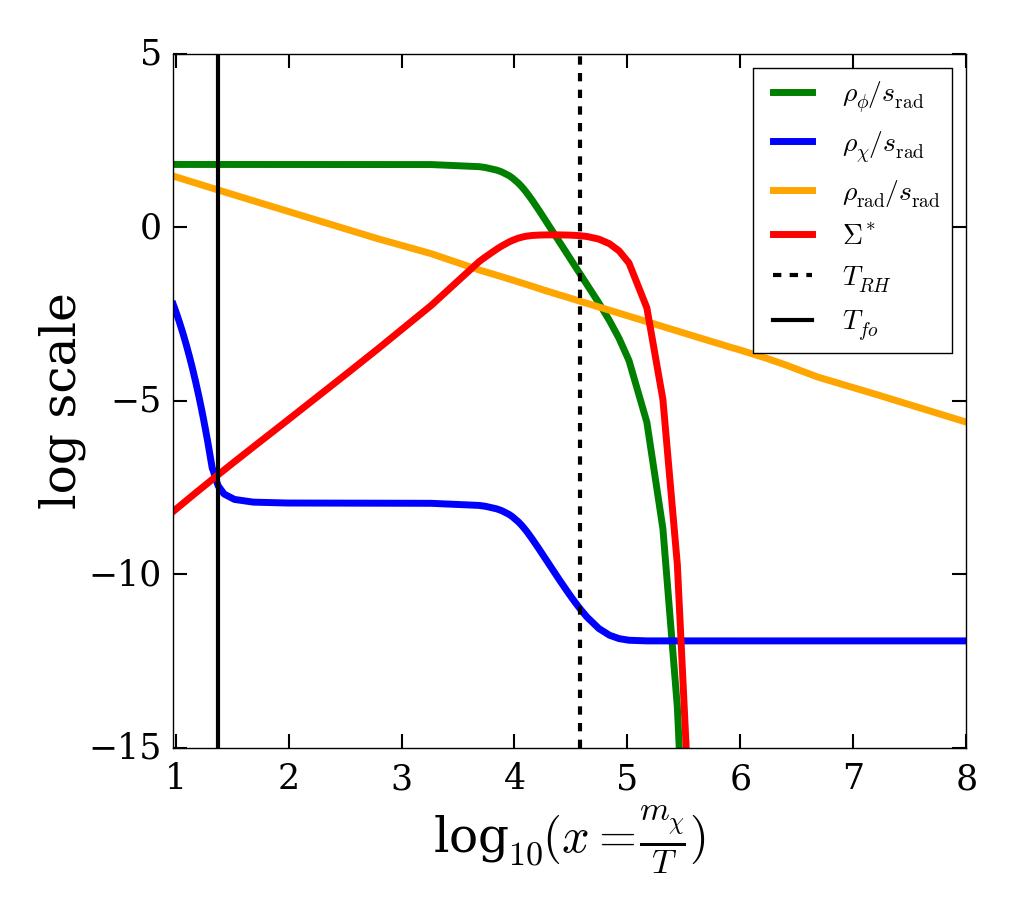}}\subfloat[$T_{RH}=10$ GeV, $\kappa_\phi=100$, $\Tinit=40$ GeV]{\includegraphics[width=0.46\columnwidth]{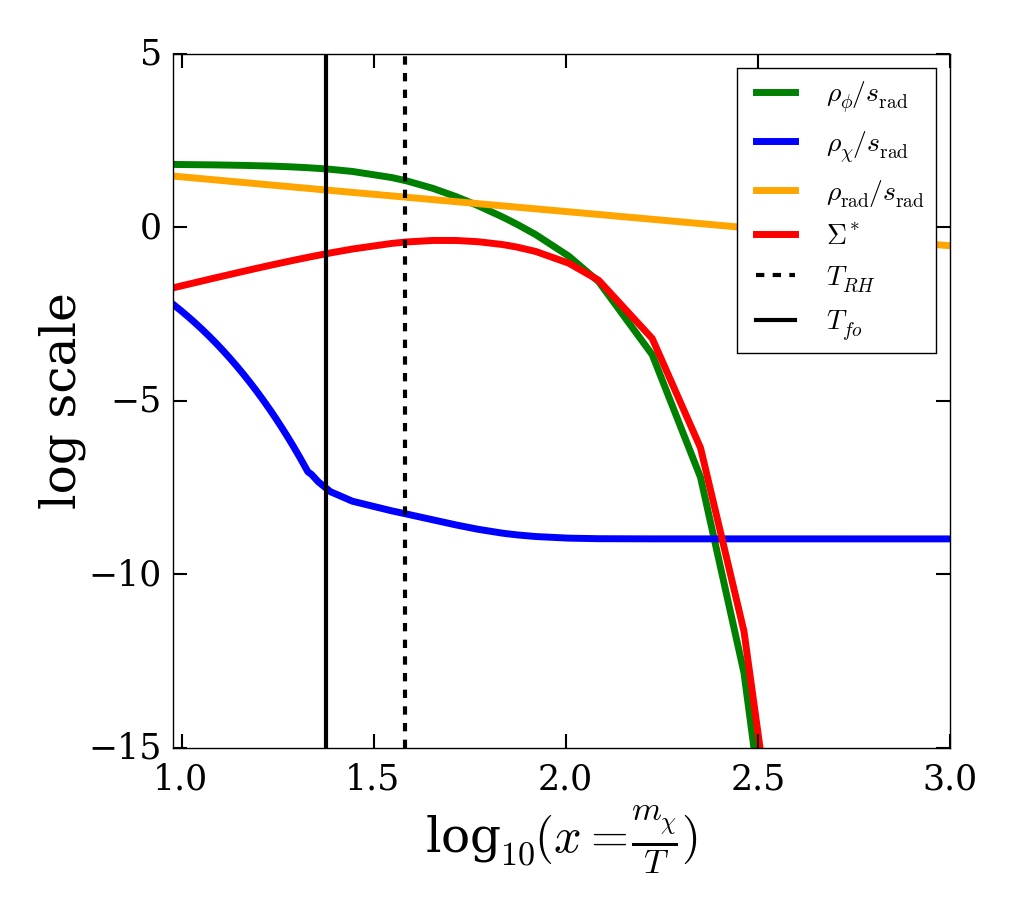}~~~~~~~~~}
\caption{\it The evolution of the scalar field, neutralino and radiation densities normalised to the radiation entropy density, and of the injection of entropy \stdstar{}, as a function of $x=m_{\chi}/T$.\label{fig:pmssm11_plotrho}}
\end{figure}

We first investigate the case where the neutralino has a relic density that is too large in the standard cosmological model, illustrated by Point A. The results of the scan over the reheating temperature \TRH\ and the initial scalar field density $\kappa_\phi$ are shown in Figure~\ref{fig:pmssm11_Tfo>TRH}, assuming
that the scalar field does not decay into SUSY particles ($\eta=0$).
We can distinguish two zones in this figure: a zone at large initial scalar field density and small reheating temperature, where the relic density is strongly reduced, and the complementary zone where the presence of the scalar field does not modify the relic density. On the one hand, the dependence on $\kappa_\phi$ of the dilution is rather clear: 
the larger $\kappa_\phi$ is, the larger \stdstar\ is initially, and the dilution is stronger. On the other hand, the value of the reheating temperature affects more the duration of the dilution than its strength. 
As illustrated in Figure~\ref{fig:pmssm11_plotrho}, when \TRH\ is small, \stdstar\ can remain at
its maximum during a large range of temperatures before its decrease due to the decay of the scalar field. 
The neutralino and scalar field densities decrease during this period with a slope $-5$, as expected
when \stdstar\ is at its maximum. For a large value of \TRH, however, the fields are diluted over a 
smaller range of temperatures and the total decrease is reduced.

Points respecting the Planck constraints, which we will refer to as \emph{accepted points}, 
lie along a thin line in the $\Log_{10}(\kappa_\phi)/\Log_{10}(T_{RH})$ plane. They follow a line of slope 
$\sim 1$ at small \TRH\  that changes slightly at $\TRH \sim 150$ MeV to a slope 1.5. 
This transition is the result of the quark/hadronic phase transition, which lowers the number of radiation degrees of freedom. 
In particular, below $T \sim 150$ MeV, pions become non-relativistic and no longer contribute to the radiation density. 
This feature is independent of the \wimp\ and scalar field properties, and is present in all the following results.

The line of accepted points becomes vertical at $\TRH \sim \Tfo$, which is to be expected when the scalar field decays completely during neutralino thermal equilibrium, as there is no possible modification of the relic density. Thus, we can derive a maximum value of the reheating temperature $\TRH \lesssim\Tfo$.
One can also note that if $\TRH < \TRH^{\rm BBN \, lim}\sim 6$ MeV, the scalar field density is too large during BBN, and the model is therefore excluded. This constraint is very general, as it is also independent of the WIMP properties, and thus applicable to any WIMP model. This limit gives us a lower bound for the reheating temperature, as well as a minimum value for the initial scalar field density $\kappa_\phi$ using $\TRH=\TRHBBN$. 
For Point A, we can deduce $\kappa_\phi \gtrsim 0.1$, but this minimum value will depend on the nature of the WIMP. 

\begin{figure*}
\centering
\subfloat[$\eta=0$]{\includegraphics[width=0.46\columnwidth ]{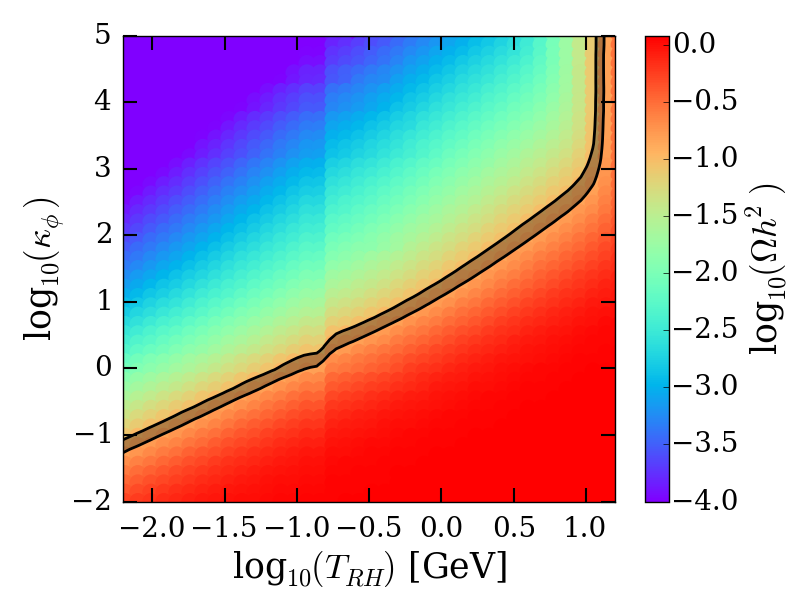}}
\subfloat[$\eta=10^{-12}$]{\includegraphics[width=0.46\columnwidth ]{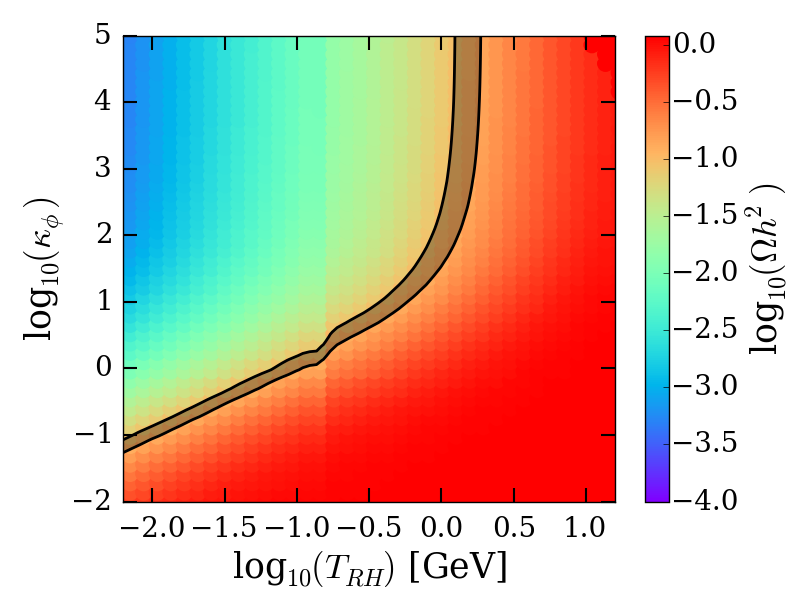}}\\
\subfloat[$\eta= 10^{-11}$]{\includegraphics[width=0.46\columnwidth ]{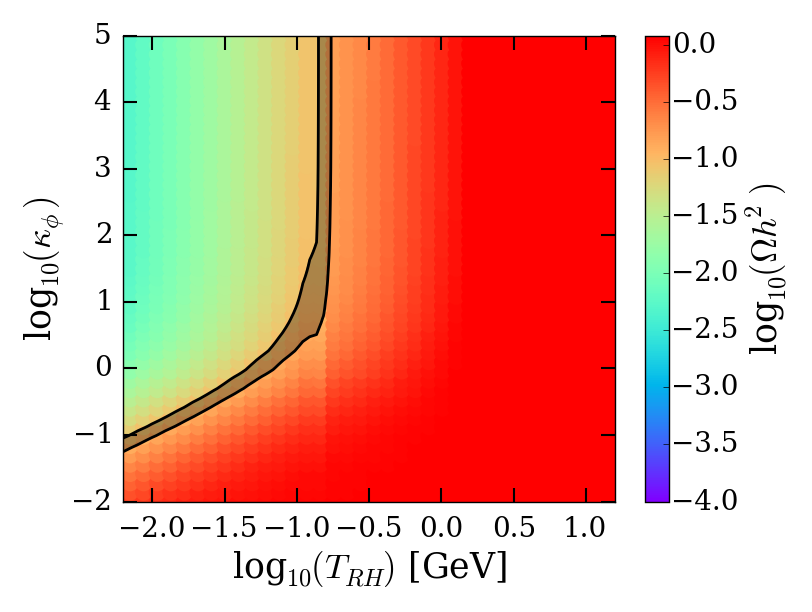}}
\subfloat[$\eta=10^{-10}$]{\includegraphics[width=0.46\columnwidth ]{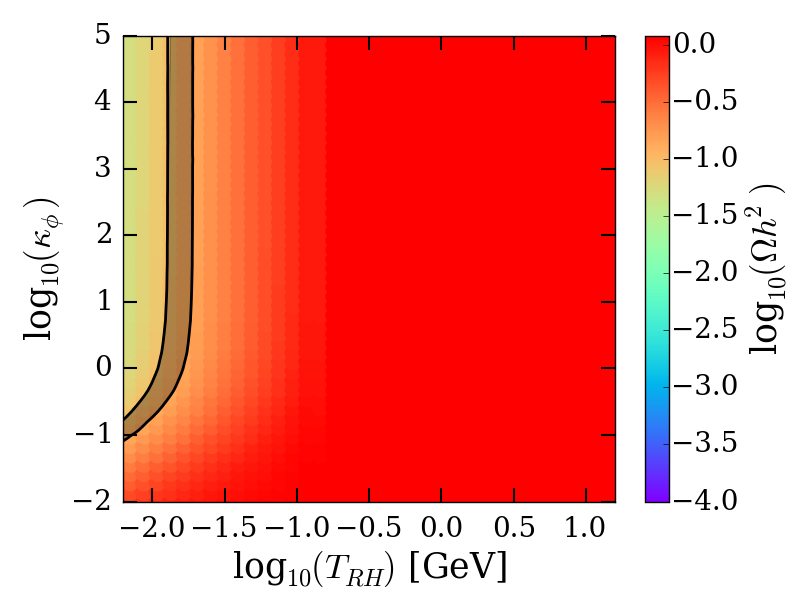}}
\caption{\it The effect of varying $\eta$ on $\Log_{10}(\Oh)$ for Point A, indicated by the colour code in the legend.\label{fig:variationeta}}
\end{figure*}

No enhancement of the relic density is possible when $\eta=0$. At small \TRH\ and large $\kappa_\phi$, 
where the scalar field density could have increased the freeze-out temperature via its relation with the Hubble {parameter}, and thereby increased the relic density, the densities are in fact already significantly reduced by dilution. 
Therefore, in order to increase the relic density, it is necessary to consider non-thermal production of the WIMP, i.e., $\eta>0$. In the case of Point A, the region of interest will be at small \TRH\ and large $\kappa_\phi$, where the relic density is strongly reduced by dilution. The scalar field decay into SUSY particles provides an additional contribution to the relic density, and the DM density measured by Planck may be reached with the appropriate value of $\eta$. We test four different values of $\eta$ in Figure~\ref{fig:variationeta}, and notice that the larger $\eta$ is, the more the line of accepted points is shifted towards small \TRH{}.

\begin{figure}
\subfloat[]{
\includegraphics[width=0.46\columnwidth ]{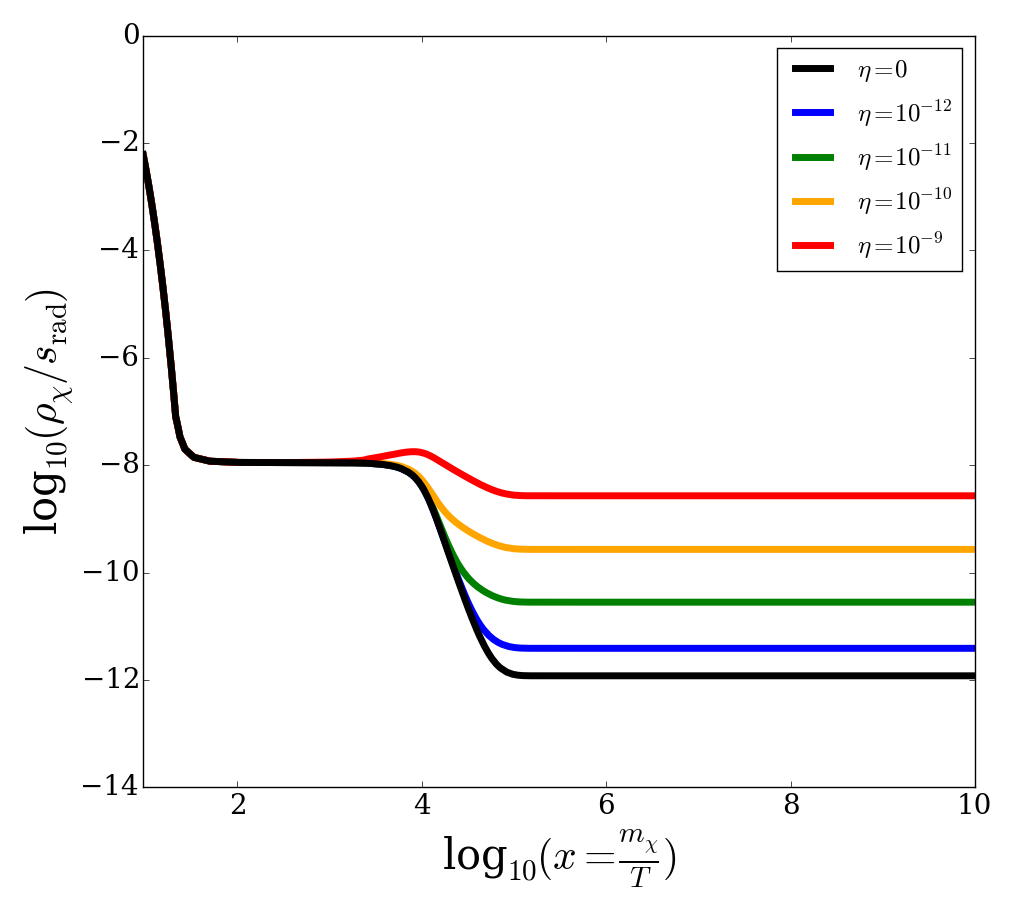}}\quad
\subfloat[]{\includegraphics[width=0.46\columnwidth]{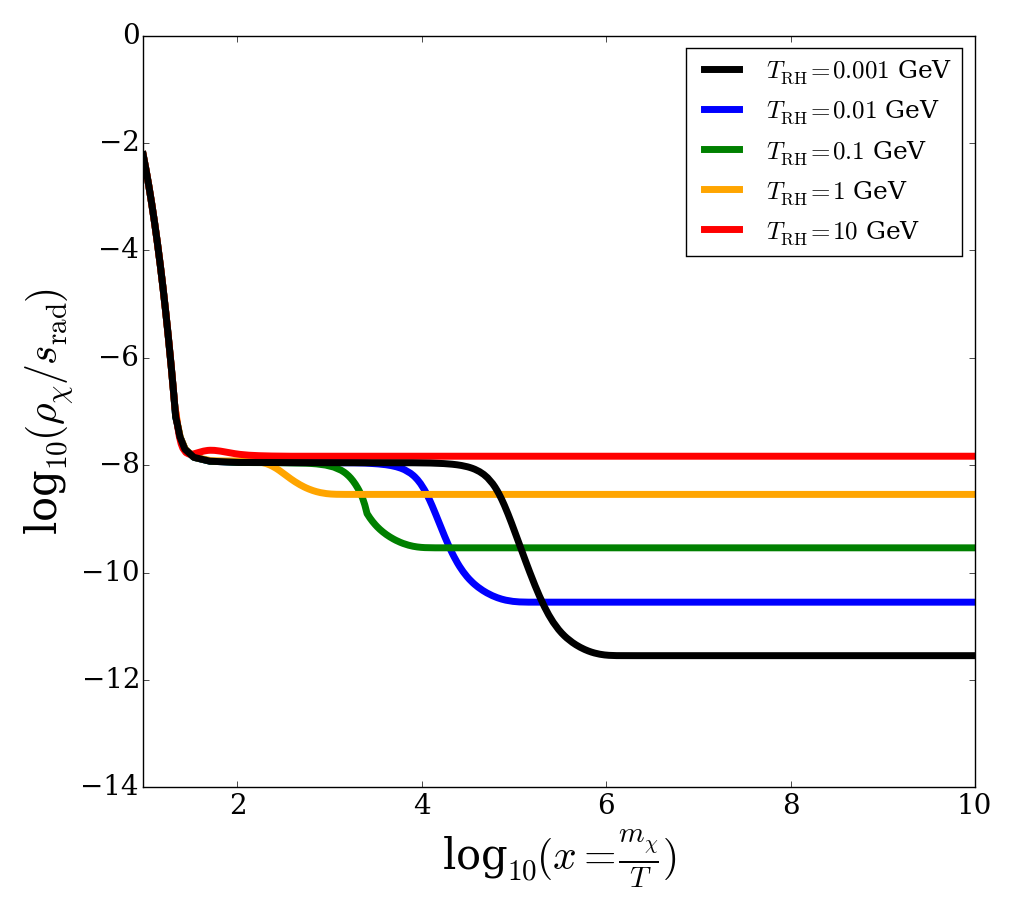}}
\caption{\it The variation of the relic density normalised to the radiation entropy density as a function of the temperature, for $\Tinit=40$ GeV and $\kappa_\phi=100$,  
when (a) varying the value of $\eta$ with fixed $\TRH=0.01$ GeV, and  (b)  varying the value of \TRH\ with fixed $\eta=10^{-11}$.\label{fig:etarho}}
\end{figure}

We observe in Figure~\ref{fig:etarho} that in the region of interest the relic density increases linearly with $\eta$ and \TRH, which explains the observed feature. Similarly to what happens with the dilution, the parameter $\eta$ impacts the strength of the non-thermal production of neutralinos, while \TRH\ impacts the time between the freeze-out and the scalar field decay, during which the relic density can benefit from this new contribution.

In the limit of large $\kappa_\phi$ and small \TRH{}, we find that the evolution of the relic density with respect to $\eta$ and \TRH\ can be approximated by:
\begin{equation}
\Oh \approx \eta \; ( \alpha \; \TRH + \beta )  \ ,
\end{equation}
where $\alpha$ and $\beta$ are numerical factors that depend, {\it a priori}, on the \wimp\ properties. When $\eta$ goes to zero, the relic density vanishes, which is expected since, in this region of the parameter space, the dilution due to the entropy injection is dominant in absence of non-thermal production. One can also note that the effects of the dilution and of the non-thermal production equilibrate in such a way that the above expression does not depend on $\kappa_\phi$.
For Point A, we find that $\alpha \approx 7.68 \times 10^{10}$ {GeV$^{-1}$} and $\beta \approx 2.62 \times 10^7$. 
This parametrisation enables us to find the value of $\eta$ required to get the correct relic density for a given reheating temperature.
On the other hand, a maximum value of $\eta$ can be calculated by considering the reheating temperature where the BBN constraints 
start excluding the model ($\TRH^{\rm lim} \approx 6\times 10^{-3}$ GeV):
\begin{equation}
\eta_{\rm Max}= \frac{{\Omega h^2}_{\rm DM}^{\rm upper\, lim}}{\alpha \TRH^{\rm lim} + \beta} \ .
\end{equation}
For our benchmark point, we calculate $\eta_{\rm Max} \approx 2.93 \times 10^{-10}$. 
Thus, in this scenario the branching ratio into SUSY particles must be very small, 
which can be traced back to our choice of a scalar field with a substantial initial density.
We note also that the variation in $\eta$ does not modify the constraints on $\kappa_\phi$ and \TRH\ that we derived in the case $\eta=0$.
Strong constraints on the scalar field parameters can therefore be derived, namely 6 MeV $\lesssim \TRH \lesssim \Tfo$, $\kappa_\phi \gtrsim 0.1$ and $\eta \lesssim 2.93 \times 10^{-10}$.

\subsubsection{Point with a small relic density}
\begin{figure*}
\centering
\subfloat[$\eta=0$]{\includegraphics[width=0.46\columnwidth]{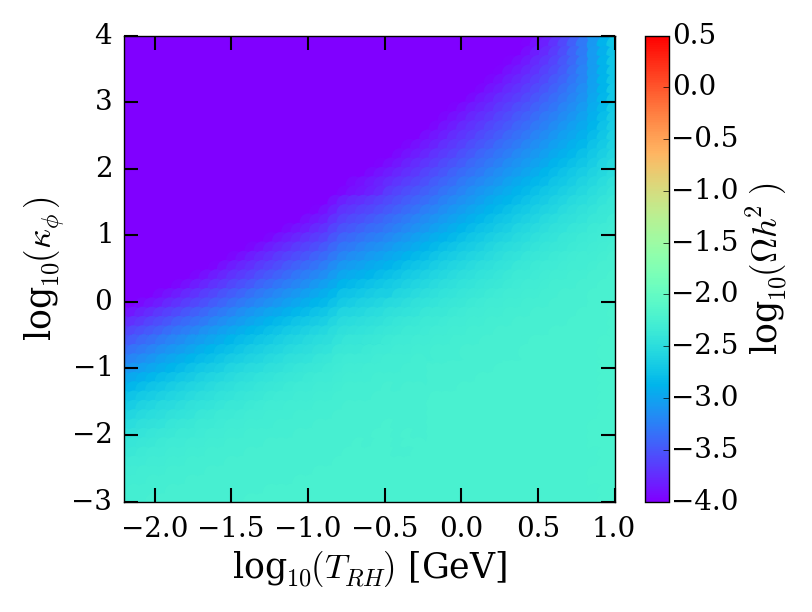}}
\subfloat[$\eta=10^{-11}$]{\includegraphics[width=0.46\columnwidth]{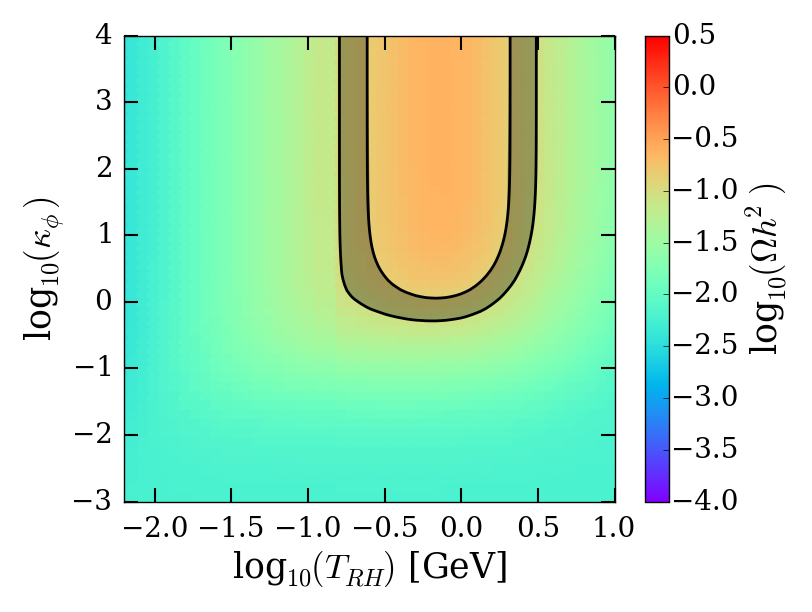}}\\
\subfloat[$\eta=10^{-10}$]{\includegraphics[width=0.46\columnwidth]{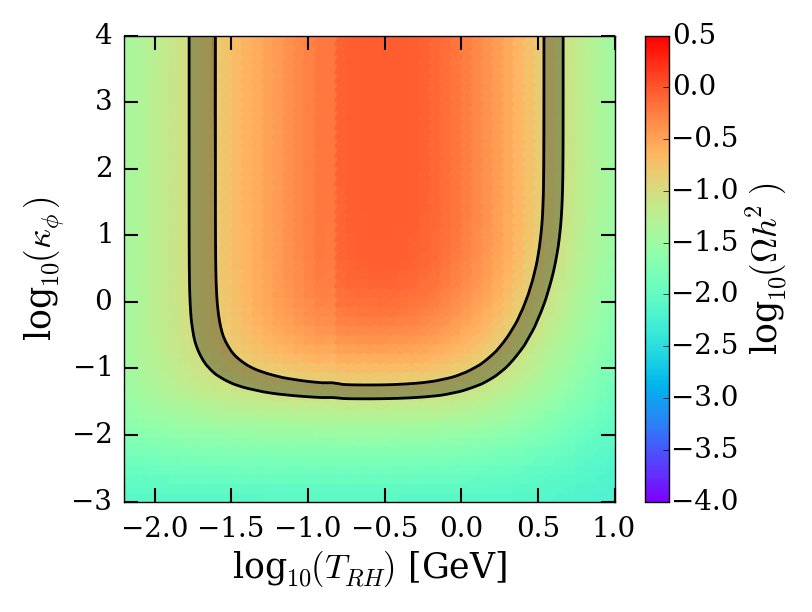}}
\subfloat[$\eta=10^{-9}$]{\includegraphics[width=0.46\columnwidth]{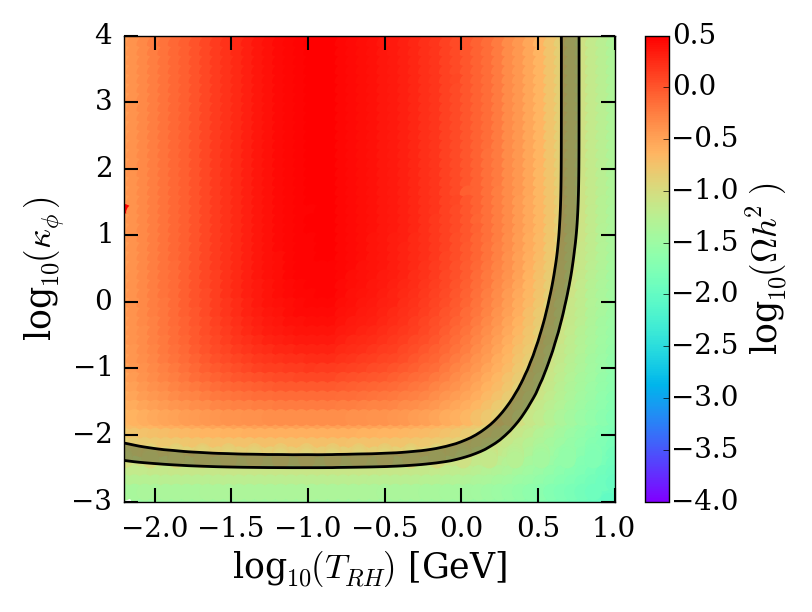}}\\
\caption{\it The effect of varying $\eta$ on $\Log_{10}(\Oh)$ for Point B, 
indicated by the colour code in the legend.\label{fig:pointBscan}}
\end{figure*}

As discussed previously, no enhancement of the relic density is possible when only entropy injection is considered. 
Therefore, one needs to allow the scalar field to decay into BSM particles. We show in Figure~\ref{fig:pointBscan}
the result of scans over \TRH\ and $\kappa_\phi$ for Point B with four different values of $\eta$. In each scenario, 
the region of accepted points forms a U shape in the $\kappa_\phi$ /\TRH\ plane. 
The vertical right limit corresponds to $\TRH \sim \Tfo$, and does not move significantly as $\eta$ increases. 
The vertical left limit, however, is shifted to the left along the \TRH\ axis and the horizontal limit is shifted
downwards towards lower values of $\kappa_\phi$. The constraints on \TRH\ that we deduced for point A hold also in this case:
$\TRHBBN \lesssim \TRH \lesssim \Tfo$. However, it is difficult to find limits on $\kappa_\phi$ and $\eta$ as stringent as the ones we found for point A.

The largest effect is in the case where the scalar field decays entirely into BSM particles and not into radiation. 
Thus, if a decay produces two SUSY particles, for example, $b=2$ and $m_{\phi}>2 m_{\chi}$, so $\eta<1/m_{\chi}$. 
In such a case, all the SUSY particles produced by the scalar field decay, starting from the neutralino freeze-out, 
constitute an overall contribution to the relic density that has to be added to the value of the relic density in the standard model,
i.e., $Y= Y_{\rm stand} +  {Y}_{\phi}^{\rm T=\Tfo}{/m_\chi}$. Therefore, one has a constraint on the scalar field density at freeze-out.

\subsubsection{pMSSM19 sample}

In the following, we study how the constraints on the scalar field depend on the \wimp\ properties disregarding
the case of a relic density that is too small, as the constraints deduced in this case already showed an explicit 
dependence on the freeze-out temperature and the relic density at freeze-out. 

\begin{figure}[t!]
\begin{minipage}[t]{0.46\textwidth}
\begin{center}
\includegraphics[width=\columnwidth]{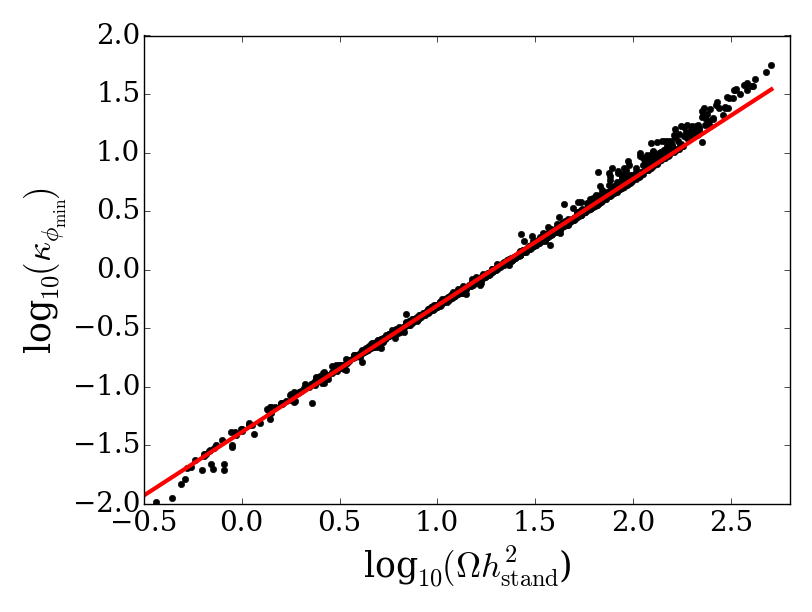}
\caption{\it The values of $\kappa_\phi$ required to reduce the relic density to the measured DM density with $\TRH=\TRHBBN$ and $\Tinit=40$ GeV as a function of the relic density calculated in the standard model of cosmology. The calculations were done for the sample of points in the pMSSM19 characterised in Table~\protect\ref{tab:pmssm}.\label{fig:find} }
\end{center}
\end{minipage}
\quad
\begin{minipage}[t]{0.46\textwidth}
\begin{center}
\includegraphics[width=\columnwidth]{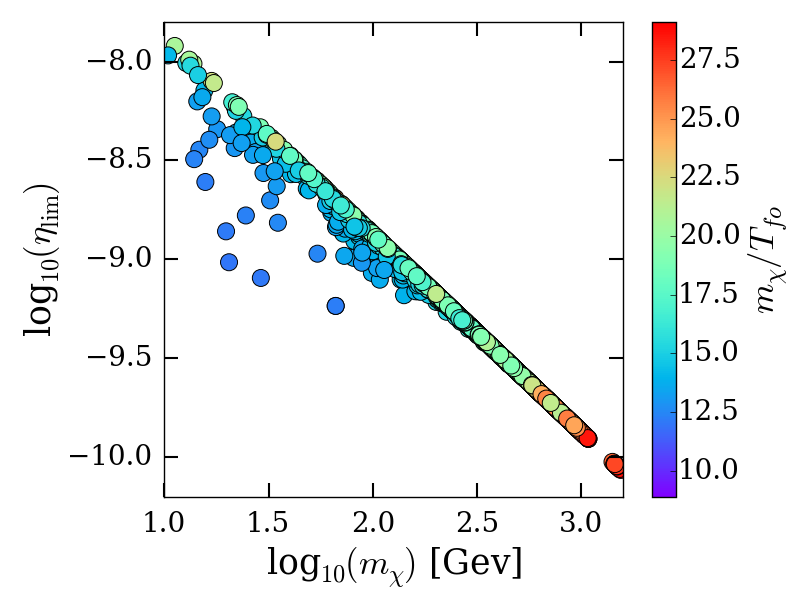}
\caption{\it The maximum value of the parameter $\eta$ for the pMSSM19 sample of points as a function of the neutralino mass. 
The values of $m_{\chi}/\Tfo$ are colour-coded as indicated in the legend. \label{fig:etalim} }
\end{center}
\end{minipage}
\end{figure}

We focus on the points in our pMSSM19 sample that have a relic density that is too large in the standard cosmological model, 
which leaves us almost exclusively with bino-like neutralinos. We calculated the values of $\kappa_\phi$ that
give the correct relic density at $\TRH=\TRHBBN$, as shown in Figure~\ref{fig:find},
and find a very good correlation between the relic density calculated in the standard model and $\kappa_{\phi_{\rm min}}$.

The points in Figure~\ref{fig:find} follow a line of slope $\sim 1$. Thus, the minimum value of the initial scalar field density 
increases with the value of the relic density in the standard model. This can be understood because the larger the relic density at freeze-out is, 
the stronger must be the dilution for a given reheating temperature. The small scatter of the points at low relic density is due to 
numerical uncertainties alone, but we note a departure from this line at large $\Oh_{\rm stand}$, 
when $\kappa_{\phi_{\rm min}}\gtrsim 1$. With a scalar field density of this order of magnitude, 
there is also a modification of the Hubble {parameter}, which advances freeze-out. This mechanism tends to 
increase the relic density, while the entropy injection decreases it. Overall, the dilution has a stronger effect, 
but a larger scalar field density is required to decrease the relic density down to the measured DM density.

Next, we calculate the maximum value of $\eta$ and find a clear dependence on the \wimp\ mass,
as seen in Figure~\ref{fig:etalim}. Indeed, the scalar field produces a fraction $b$ of SUSY particles, 
which contributes as $m_{\chi} \times b$ to the \wimp\ mass density. Therefore, the larger $m_{\chi}$ is,
the more the relic density will be increased for a given value of $\eta$, and the smaller will be the maximum value of $\eta$. 
At first approximation, the maximum value of $\eta$ is inversely proportional to the \wimp\ mass. However, 
another mechanism is at play: for the same neutralino mass, the larger $\Tfo$ is, the larger the neutralino density at the freeze-out temperature is, and thus the smaller $\eta$ must be in order to reach the correct relic density. 
As $T_{\rm fo_{\rm stand}} \approx m_{\chi}/20$, we can express a linear relation between $\eta_{\rm lim}$ 
and $m_{\chi}$. However, as shown in Figure~\ref{fig:etalim}, when $\Tfo$ departs from this approximation towards larger values, the second mechanism becomes more important, and we see a departure from the linear relation between $m_{\chi}$ and $\eta_{\rm lim}$. This happens for neutralino masses smaller than $\sim 100$~GeV in our sample of points.
In any case, $\eta$ must be very small, of the order of $\sim 10^{-10}$ -- $10^{-9}$.

\subsection{Quintessence}

We now turn to the study of the quintessence model. This scenario only has the power to increase the relic density by advancing freeze-out. Therefore, we disregard the case of a standard relic density that is too large.

\subsubsection{Point with a small relic density}

\begin{figure}[t!]
\centering
\subfloat[$n_2=3$]{
\includegraphics[width=0.46\columnwidth]{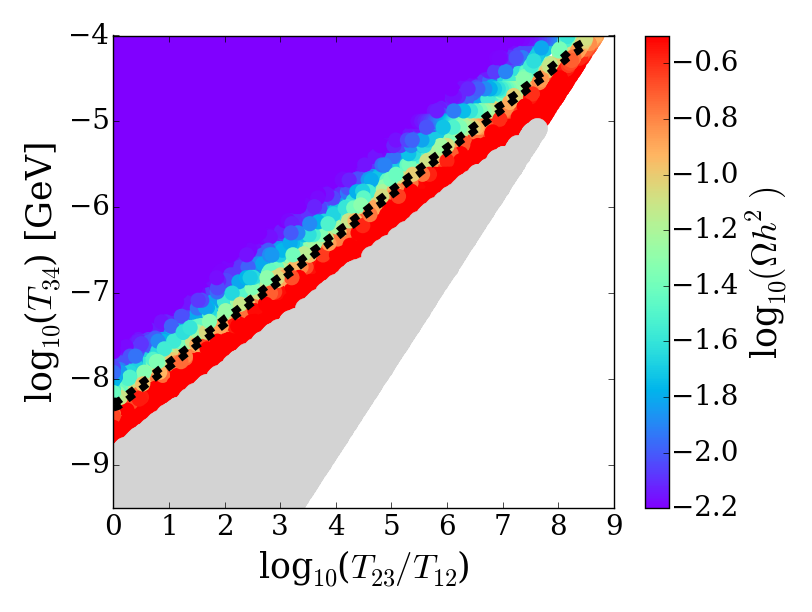}}
\subfloat[$n_2=6$]{
\includegraphics[width=0.46\columnwidth]{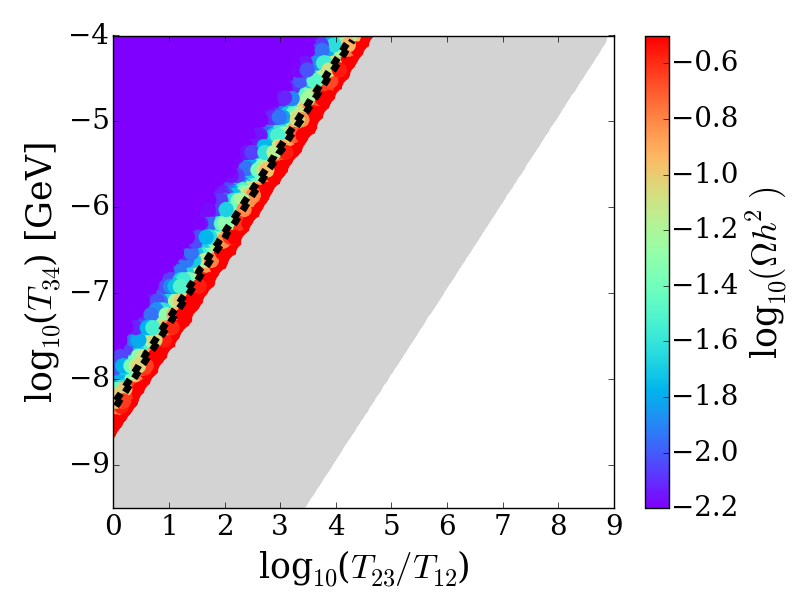}}
\caption{\it The value of $\Log_{10}(\Oh)$, colour-coded as indicated in the legend, in the $T_{34}$, $T_{23}/T_{12}$ parameter plane of the quintessence model. The accepted parameter sets lie between the two dashed lines, the grey region is excluded by BBN and the white region is not accessible in this model.\label{fig:quintessence}}
\end{figure}

We have scanned over the three temperature parameters such that $T_0<T_{12}<T_{23}<T_{34}$ 
with $T_0=2 \times 10^{-13}$ GeV, the temperature of the CMB at present time.
We performed the scans for the two extreme values of the slope in zone 2 of Figure~\ref{fig:model},
namely $n_2=3$ and $n_2=6$. We have calculated the relic density of our benchmark CMSSM point for each set of quintessence parameters, and show the results in Figure~\ref{fig:quintessence}.

The relevant parameters are $T_{34}$ and the ratio $T_{23}/T_{12}$. The smaller $T_{34}$ is,
and the greater $T_{23}/T_{12}$ is, the larger is the relic density. This can easily be understood as
the larger the scalar field density is around freeze-out, the larger will be the increase of the relic density, and a small value of $T_{34}$ and a large difference between $T_{12}$ and $T_{23}$ helps in
obtaining a large scalar field density at large temperatures. In the case $n_2=3$, the accepted parameter
sets follow a line of slope $\sim 0.5$, and we find a limit at $T_{23}/T_{12} \sim 6 \times 10^8$ and $T_{34}\sim 10^{-4}$ GeV, where the line reaches the limiting case  $T_{34}=T_{23}$. A minimum value of $T_{34}$ can be found when $T_{12}=T_{23}$, where we find $T_{34}\gtrsim 2 \times 10^{-9}$ GeV. In the case where $n_2=6$, the same minimal value can be found. However, the accepted parameter sets follow a line of slope 1, parallel to the limit $T_{23}=T_{34}$. There are, therefore, no maximum values for the temperature parameters. 

In both cases, we note also that the accepted parameter sets are very close to the limit imposed by BBN, which mainly depends on the density of the scalar field at a temperature $T \sim 1$ MeV.

\begin{figure}[t!]
\centering
\includegraphics[width=0.5\columnwidth]{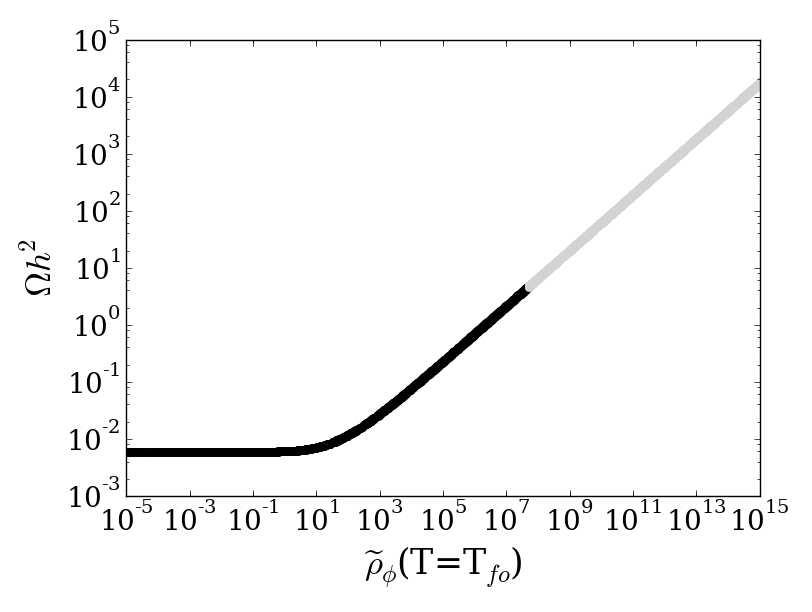}
\caption{\it The increase in the relic density for Point B as a function of the ratio of the scalar density and the radiation density at 1 MeV. The grey region is excluded by BBN.\label{fig:rhophi1MeV}}
\end{figure}

When $T_{34}$ is smaller than 1 MeV, which must be the case for values of $n_2$ close to 3, it is possible to find simpler constraints on the scalar field properties. In this case, freeze out and BBN both occur during phase 4 of the scalar field evolution in the model. The scalar field density can thus be specified simply by its value at freeze-out, and determined at other temperatures according to the slope $n_4=6$. We can therefore disregard what happens in phases 1, 2 and 3.
We show in Figure~\ref{fig:rhophi1MeV} the evolution of the relic density for Point B with the ratio of the scalar field density to the radiation density at freeze-out, $\widetilde{\rho}_{\phi}=\frac{\rho_\phi}{\rho_{rad}}(T=\Tfo)$ when we consider only phase 4 of the model.

The scalar field starts having an effect on the relic density when its density is comparable to the radiation density at freeze-out. 
The Hubble {parameter} is thus significantly modified and freeze-out is advanced. The relic density then increases with a slope $\sim 0.48$. In addition, we note that points are excluded by BBN 
if $\dfrac{\rho_{\phi}}{\rho_{rad}}$(T=\Tfo)$\gtrsim 10^{8}$, which corresponds to {$\dfrac{\rho_{\phi}}{\rho_{rad}}$(1 MeV)$\gtrsim 1$}.

\subsubsection{pMSSM19 sample}
\begin{figure}[t!]
\centering
\includegraphics[width=0.5\columnwidth]{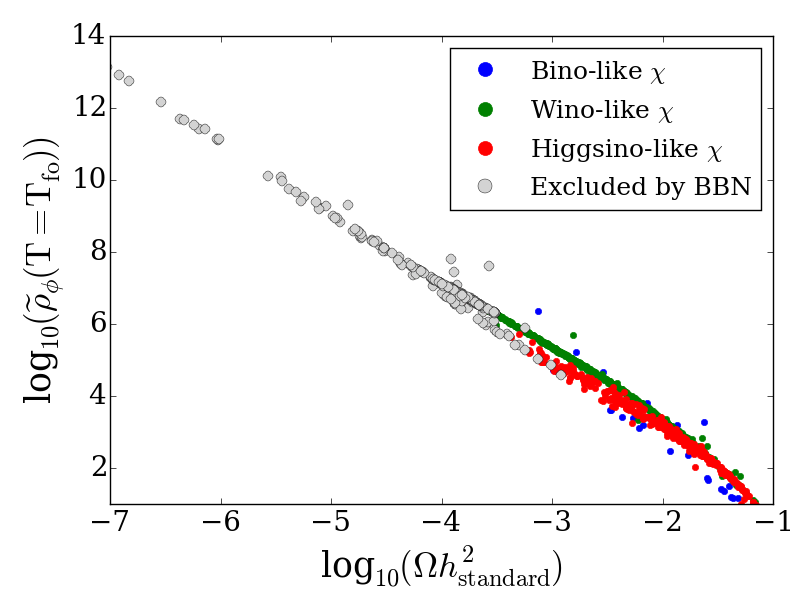}
\caption{\it The value of the scalar field density at freeze-out that is required to increase the relic density up to the observed DM density for our sample of pMSSM19 points. The neutralino mass is shown in colour and parameter sets excluded by BBN are shown in grey.\label{fig:rhoOh2_1}}
\end{figure}

In addition, we have calculated the value of $\rhotphi(T=\Tfo)$ required to obtain the correct relic density in our sample of pMSSM19 points. 
The result is presented in Figure~\ref{fig:rhoOh2_1}, which shows the dependence of $\rhotphi(T=\Tfo)$ on the standard relic density. 

In a first approximation, $\rhotphi(T=\Tfo)$ scales as a power of the standard relic density, with an exponent $\sim -2$. The smaller the standard relic density is, the larger the scalar field density must be around freeze-out in order to increase the relic density up to the DM density. The exponent $-2$ can be understood from a simple calculation. Freeze-out occurs when the annihilation rate equals the expansion rate, in the standard cosmological model:
\begin{equation}
n_{eq}(\Tfo^{stand}) \sigmav_{T_{fo_{stand}}} \sim H \sim H_{0} \rho_{rad}^{1/2}(T=T_{fo_{stand}}) \ ,
\label{eq:e1}
\end{equation}  
with $H_0=\sqrt{8\pi / 3 M_{p}^{2}}$. The comoving neutralino density $Y_{stand}$ can then be expressed as:
\begin{equation}
Y_{stand}=\frac{n_{eq}(T_{fo_{stand}})}{s_{rad}(T_{fo_{stand}})} \ ,
\label{eq:e2}
\end{equation} 
which can be re-expressed using Eq. (\ref{eq:e1}) as
\begin{align}
Y_{stand}&=\frac{H_{0} \rho_{rad}^{1/2}(T_{fo_{stand}})}{\sigmav_{T_{fo_{stand}}} s_{rad}(T_{fo_{stand}})} \ .
\label{eq:e3}
\end{align}
When the scalar field density is very large in the quintessence model, compared to the radiation density, we obtain similar equations:
\begin{equation}
n_{eq}(\Tfo) \sigmav_{T=\Tfo} \sim H \sim H_{0} \rhophi^{1/2}(T=\Tfo)= H_{0} \rhophi^{1/2}(T=T_{fo_{stand}}) \times \left(\frac{\Tfo}{T_{fo_{stand}}} \right)^3 \ ,
\label{eq:e4}
\end{equation}  
and
\begin{equation}
Y=\frac{n_{eq}(\Tfo)}{s_{rad}(\Tfo)} \ ,
\label{eq:e5}
\end{equation} 
where we have used in Eq. (\ref{eq:e4}) the fact that the scalar field density evolves as $T^{n_4}$ with $n_4=6$. 
The relic comoving density $Y$ in this scenario can then be re-written using Eq. (\ref{eq:e4}) as:
\begin{equation}
Y=\frac{H_{0} \rhophi^{1/2}(T=T_{fo_{stand}}) \times \left(\frac{\Tfo}{T_{fo_{stand}}} \right)^3}{\sigmav_{\Tfo} s_{rad}(\Tfo)}=\frac{H_{0} \rhophi^{1/2}(T=T_{fo_{stand}})}{\sigmav_{\Tfo} s_{rad}( T_{fo_{stand}})} \ .
\label{eq:e6}
\end{equation}
Finally, we can combine Eqs. (\ref{eq:e6}) and (\ref{eq:e3}) to obtain:
\begin{equation}
Y= Y_{stand} \frac{\sigmav_{T_{fo_{stand}}}}{\sigmav_{\Tfo}} \frac{\rhophi^{1/2}(T=T_{fo_{stand}})}{\rho_{rad}^{1/2}} \ .
\end{equation}
This gives us the ratio between the scalar field density and the radiation density at the standard freeze-out temperature that is required to increase the relic density to the measured dark matter density:
\begin{align}
\begin{aligned}
\rhotphi(T_{fo_{stand}})&= \left(\frac{Y}{Y_{stand}} \right)^2 \times \left(\frac{\sigmav_{\Tfo}}{\sigmav_{T_{fo_{stand}}}} \right)^2\\
&= \left(\frac{\Oh_{DM}}{\Oh_{stand}} \right)^2 \times \left(\frac{Y(T=\Tfo)/Y(T=present)}{Y_{stand}(T=T_{fo_{stand}})/Y_{stand}(T=present)} \right)^2 \times \left(\frac{\sigmav_{\Tfo}}{\sigmav_{T_{fo_{stand}}}} \right)^2 \ .
\end{aligned}
\end{align}
We retrieve here the slope $-2$. We note, however, that this particular value appears only because $n_4=6$,
and thus depends on the quintessence model. Residual annihilations occurring after freeze-out are taken into account by the factor 
\[ \xi= \left(\dfrac{Y(T=\Tfo)/Y(T=present)}{Y_{stand}(T=T_{fo_{stand}})/Y_{stand}(T=present)} \right)^2\,,\]
which takes a value~$\sim 10$~in our sample of pMSSM19 points. It was indeed already noted in \cite{DEramo:2017gpl} that the residual annihilations, so-called \emph{relentless annihilations}, can be particularly important when $H \propto T^{2+ \frac{n}{2}}$, with $n \geq 2$. In the case of the quintessence model, $n=2$, which corresponds well to this regime.
The value of $\xi$ is model-dependent, however, and we show in Figure~\ref{fig:rhoOh2_1} that wino-like neutralinos, for instance, require a larger scalar field density than higgsino-like neutralinos.

Finally, we note that for neutralinos with a standard relic density $\lesssim 3\times 10^{-4}$, the scalar field density is too large at 1 MeV and our scenario is ruled out by BBN.

\section{Conclusions}
\label{sec:conclusion}
The cosmological density of cold dark matter is now known with good accuracy,
thanks to measurements by Planck and other cosmological and astrophysical
observations. We have studied in this paper how this knowledge could be used to
constrain possible non-standard evolution of the early Universe in specific dark
matter scenarios. An optimist might assume that laboratory experiments would establish
the parameters of some scenario for physics beyond the Standard Model
sufficiently well for a discrepancy to be established between the cosmological measurements
and model calculations in standard radiation-dominated cosmology. More conservatively, 
the combination of observations and model calculations could be used to
constrain a combination of model parameters and early-Universe scenarios.

As examples of non-standard evolution in the early Universe before Big Bang Nucleosynthesis, 
we have considered scenarios in which a scalar field decays into some combination of
Standard Model and other particles, and quintessence models with various classes of
effective potential. Our calculations were illustrated using various supersymmetric models 
in which a calculation of the cold dark matter density assuming a conventional radiation-dominated 
early Universe would yield a density that is either larger or smaller than the observed density.
The measured cold dark matter density could be used in the case of a decaying scalar field
to constrain the initial density of the scalar field, the reheating temperature after it decays, and
the branching ratio for its decays into particles beyond the Standard Model.
In the case of a quintessence model, the cold dark matter density could be used to
constrain the evolution with temperature in the early Universe of the quintessence field.

Our results exemplify the idea that measurements by laboratory experiments 
could be used, in the context of a specific model for physics beyond the Standard Model,
to constrain aspects of the physics controlling the evolution of the early Universe that would
otherwise be invisible and inaccessible. In this way, collider and other laboratory experiments
could serve as powerful telescopes, using dark matter particles as a novel type of messenger
particle able to provide information about the early Universe that photons and neutrinos cannot provide.

\section*{Acknowledgements}
The work of JE was supported in part by the United Kingdom STFC Grant ST/P000258/1, and in part
by the Estonian Research Council via a Mobilitas Pluss grant.


\bibliographystyle{JHEP}
\bibliography{biblio}

\end{document}